\documentclass[twocolumn,showpacs,preprintnumbers,amsmath,amssymb]{revtex4}
\usepackage{graphicx}% Include figure files
\usepackage{dcolumn}% Align table columns on decimal point
\usepackage{bm}% bold math

\begin{document}
\title{Dynamical response of the nuclear ``pasta'' in neutron star crusts}
\author{C.J. Horowitz\footnote{e-mail: horowit@indiana.edu} 
        and M.A. P\'{e}rez-Garc\'{\i }a} 
\affiliation{Nuclear Theory Center and Department of Physics, 
             Indiana University, Bloomington, IN 47405}
\author{D. K. Berry\footnote { e-mail: dkberry@indiana.edu}}
\affiliation{University Information Technology Services,
             Indiana University, Bloomington, IN 47408}
\author{J. Piekarewicz\footnote{e-mail: jorgep@csit.fsu.edu}}
\affiliation{Department of Physics,
         Florida State University, Tallahassee, FL 32306}

\date{\today} 
\begin{abstract}
The nuclear pasta --- a novel state of matter having nucleons arranged
in a variety of complex shapes --- is expected to be found in the crust of neutron stars and in core-collapse supernovae at subnuclear densities of about $10^{14}$~g/cm$^3$.  Due to frustration, a phenomenon that emerges from the competition between short-range nuclear attraction and long-range Coulomb repulsion, the nuclear pasta displays a preponderance of unique low-energy excitations.  These excitations could have a strong impact on many transport properties, such as neutrino propagation through stellar environments. The excitation spectrum of the nuclear pasta is computed via a molecular-dynamics simulation involving up to 100,000 nucleons. The dynamic response of the pasta displays a classical plasma oscillation in the 1-2 MeV region.  In addition, substantial strength is found at low energies. Yet this low-energy strength is missing from a simple ion model containing a single-representative heavy nucleus. The low-energy strength observed in the dynamic response of the pasta is likely to be a density wave involving the internal degrees of freedom of the clusters.  
\end{abstract}
\maketitle 

Baryonic matter is organized as a result of short-range nuclear attraction and long-range Coulomb repulsion.  Often the corresponding nuclear and atomic length scales are well separated, so nucleons bind into atomic nuclei that are themselves segregated into a crystal lattice. However, at the enormous densities present in astrophysical objects --- densities that exceed that of ordinary
matter by 14 orders of magnitude --- these length scales become comparable and complex new phenomena emerge.  Complexity arises because it is impossible for the constituents to be simultaneously correlated from nuclear attraction and anti-correlated from Coulomb repulsion.  Competition among these interactions plays a fundamental role in the organization of matter and results in {\it Coulomb frustration}.  Frustration --- a ubiquitous behavior in complex systems ranging from magnetism to protein folding to neural networks --- develops from the inability of a system to simultaneously satisfy all of its elementary interactions.  For example, the Ising antiferromagnet on a triangular lattice is frustrated because not all of the nearest neighbor spins can be anti-parallel to each other.   Frustrated systems have unusual dynamics due to the preponderance of low-energy excitations~\cite{frustrated}.

At subnuclear densities of about $10^{14}$ g/cm$^3$ (normal nuclear matter saturation density is $2.5\times10^{14}$ g/cm$^3$) Coulomb frustration is expected to promote the development of complex shapes.  These shapes follow from the competition between surface tension and Coulomb energies. While surface tension favors spherical shapes, Coulomb interactions often favor non-spherical configurations. Therefore, a variety of complex structures with a diversity of shapes --- such as spheres, cylinders, plates {\it etc.}--- have been predicted.  The many phases of nuclear matter displaying this variety of shapes are known collectively as {\it nuclear pasta}. The complex dynamics of the nuclear pasta is of relevance to the structure of the inner crust of neutron stars and to the dynamics of core-collapse supernovae~\cite{pasta}.

There have been several calculations of the ground-state shapes of the
nuclear pasta~\cite{pasta_shapes}. However, to our knowledge there
have been almost no calculations of the dynamical properties of the
nuclear pasta.   Some dynamical aspects of nuclei in the inner crust of neutron stars were considered by Magierski and Bulgac \cite{magierski} while Khan, Sandulescu, and Van~Giai have calculated the excitation spectrum of pasta in a random phase approximation (RPA)~\cite{vangi}.  They find a low energy collective oscillation of the neutron rich skin of the pasta.  We note, however, that their use of a spherical Wigner-Seitz unit cell may be a serious drawback in the case of long wavelength collective modes that can extend across many unit cells. Here we are interested in the low-momentum (long wavelength) dynamic response of the pasta, that we compute from a semi-classical Molecular-Dynamics simulation containing up to 100,000 nucleons. As the response of the nuclear pasta at low momentum transfers is dominated by heavy clusters, their thermal de~Broglie wavelength is very small compared to the inter-cluster separation. This fact motivates our semi-classical approach.

The nuclear pasta is a unique hybrid state consisting of both atomic
and nuclear matter. Therefore, the excitation spectrum involves both
atomic and nuclear modes that may occur at similar energies. This
allows for a unique mixing between atomic and nuclear excitations. For
example, a dense system of charged particles displays plasma oscillations while nuclei exhibit collective density oscillations known as giant resonances. Such novel excitation spectrum may strongly impact on a variety of transport properties, such as thermal conductivity, viscosity, diffusion, and opacity. These are all important for many neutron-star observables and could influence the dynamics of core-collapse supernovae. Indeed, in core-collapse
supernovae --- the giant explosions of massive stars --- 99\% of the
energy of the explosion is radiated in the form of neutrinos.  In a
previous work~\cite{pasta1,pasta2} we have computed the {\it static
structure factor} of the pasta and the resulting neutrino mean free
path that is dominated by coherent scattering from the various pasta
shapes. Here we extend our previous work to study the excitation
spectrum of the pasta by computing its {\it dynamical response}.  This
may influence the energy equilibration between matter and $\mu$ and
$\tau$ neutrinos.  We note that many complex fluids, such as polymers, colloids, water-surfactant-oil solutions, microemulsions, and liquid crystals, display similar complex shapes (see for example~\cite{complexfluids}). Neutron and X-ray scattering from these systems probes these complex shapes in a manner that is analogous to neutrino scattering from nuclear pasta.

The nuclear pasta is described through a simple semi-classical model
that we have used earlier to compute its static structure 
factor~\cite{pasta1,pasta2}.  Here we are interested in computing
its dynamical response to neutrino scattering at low momentum
transfers.  We model the nuclear pasta as a charge-neutral system of
neutrons, protons, and electrons.  At the relevant densities and
temperatures of our simulations, the electrons form a degenerate,
relativistic Fermi gas that is not modeled explicitly. Rather, the
electrons modify the Coulomb interaction between the protons through a
screening length $\lambda$.  Neutrons and protons are described by the
following semi-classical Hamiltonian: $H=K+\sum_{i<j} v_{ij}$, where
$K$ represents the kinetic energy for nucleons of mass $m$ and the 
two-body potential is given by
\begin{equation}
  v_{ij}(r)= a {\rm e}^{-r^2/\Lambda} 
           + b_{ij}{\rm e}^{-r^2/2\Lambda} 
           + e_i e_j \frac{{\rm e}^{-r/\lambda}}{r}\;.
\label{v}
\end{equation}
Here $e_i$ is the electric charge of the $i_{}$th nucleon and the
parameters of the model are: $a\!=\!110$ MeV, $b_{ij}\!=\!-2$ MeV for
the interaction between either two neutrons or two protons,
$b_{ij}\!=\!-50$ MeV for the interaction between a neutron and a proton, and $\Lambda\!=\!1.25$~fm$^2$. These parameters have been fitted so that molecular dynamics simulations at a temperature of 1 MeV reproduce a saturation density $\rho=0.16$ fm$^{-3}$ and a binding energy near -16 MeV~\cite{pasta1}.  These are appropriate values for symmetric nuclear matter near zero temperature.  This is enough to reproduce the main features of pasta formation, that matter clumps into clusters of appropriate density.  However, the clusters can not be too large because of the Coulomb interaction.  Note that our semiclassical model is not directly applicable at zero temperature.  We keep the value of the screening length fixed at $\lambda\!=\!10$~fm in order to compare with our earlier
calculations. (This value is slightly smaller than the Thomas-Fermi
screening length of the electron gas.) Watanabe and collaborators have
computed static properties of the nuclear pasta by performing
Quantum-Molecular-Dynamics simulations with a more complicated
interaction~\cite{watanabe}. Our aim here is to employ a ``minimal
model'' that, by incorporating both nuclear saturation and Coulomb
repulsion, may capture the essential features of frustration
in a transparent fashion.

The differential cross section for neutrino scattering from the 
nuclear pasta may be written as follows~\cite{pasta1}:
\begin{equation}
 \frac{d\sigma}{d\Omega dE}=\frac{G_F^2E_\nu^2}{4\pi^2} 
 \Bigl[c_a^2(3-x )S_A(q,\omega) + c_v^2(1+x)S_{V}(q,\omega)\Bigr],
 \label{sigma}
\end{equation}
where $G_F$ is the Fermi constant, $E_\nu$ is the neutrino energy,
$x=\cos\theta$ (with $\theta$ the scattering angle), and the weak
vector charge of a nucleon is $c_v\!=\!-1/2$ for a neutron and
$c_v\!\approx\!0$ for a proton. Further, the dynamic response is
probed at a momentum transfer $q$ and at an energy transfer
$\omega$. The axial term involving $c_a\!=\!\pm 1.26/2$ and the
dynamical spin response $S_A(q,w)$ will be discussed in a later work.
Here we focus exclusively on the vector (density) response
$S(q,\omega)\!\equiv\!S_{V}(q,\omega)$ that should be greatly enhanced
by coherent effects.

The dynamical response of the system to a density perturbation
is given by
%%%
\begin{equation}
  S(q,\omega)=\frac{1}{\pi} \int_0^{T_{\rm max}} 
               S(q,t) \cos(\omega t)\, dt \;.
\label{sqw}
\end{equation}
%%%
Here $S(q,t)$ represents the ensemble average of the density-density 
correlation function that is computed as the following time average:  
%%%
\begin{equation}
  S(q,t)=\frac{1}{N}\frac{1}{T_{\rm ave}}\int_0^{T_{\rm ave}} 
         \rho({\bf q},t+s)\rho(-{\bf q},s) ds \;.
 \label{sqt}
\end{equation}
%%%
In the above expression $N$ is the number of neutrons in the system 
and we discuss choices for $T_{\rm max}$ and $T_{\rm ave}$ below. 
Note that in order to improve statistics, an angle average of 
Eq.~(\ref{sqt}) over the direction of ${\bf q}$ has been performed. 
Finally, the one-body neutron density is given by $\rho({\bf q},t)=\sum_{i=1}^N 
 \exp[{i{\bf q}\cdot {\bf r}_i(t)}]$
with ${\bf r}_i(t)$ the position of the $i_{}$th neutron at time 
$t$. Note that because $c_v\!\approx\!0$ for protons, the sum over
$i$ runs only over neutrons. Further, the static 
structure factor computed in Ref.~\cite{pasta2} is easily recovered 
from $S(q)\equiv S(q,t\!=\!0)=\int_0^\infty S(q,\omega) d\omega$. 

$S(q,\omega)$ is now computed for conditions studied in 
an earlier publication~\cite{pasta2}. These include a fixed temperature
of $T\!=\!1$~MeV, a proton-to-baryon fraction of $Y_p\!=\!0.2$, and
baryon densities of $\rho=0.01$~fm$^{-3}$ and $\rho=0.05$ fm$^{-3}$
(these represent about $1/15$ and $1/3$ of normal nuclear
density). Note that during core-collapse supernova, the proton
fraction starts near $1/2$ and drops to about $0.1$ due to electron
capture, so $Y_p\!=\!0.2$ is a representative value.
To fit a 10 MeV neutrino with a 120 fm wavelength into the simulation volume, we must include up to 100,000 nucleons in our molecular-dynamics simulations.  We start the first simulation by distributing 80,000 neutrons and 20,000 protons at random within the simulation volume and with their velocities selected according to a Maxwell-Boltzmann distribution at a temperature of $T\!=\!1$~MeV. At each time step of $\Delta t\!=\!1\!-\!2$~fm/c, Newton's equations of motion are integrated using a standard velocity Verlet algorithm~\cite{verlet}. This time step typically conserves energy to at least one part in $10^4$.  The system is thermalized by evolving for a time (28,000 fm/c), during which the velocities are periodically rescaled to maintain the temperature fixed at $T\!=\!1$~MeV. To calculate $S(q,\omega)$ the system is evolved, without any velocity rescaling\cite{footnote}, for a further time of $388,000$~fm/c$=T_{ave}+T_{max}$ [see Eq.~(\ref{sqt}) and below for $T_{max}$] during which the spatial configurations of all 100,000 nucleons are written to disk every 20~fm/c --- for a grand total of 19,400 configurations.    The simulations were performed on four special purpose accelerated MDGRAPE-2 boards~\cite{MDGRAPE,pasta2} with a combined performance of roughly 500 times that of a single conventional CPU.

%%%%%%%%%%%%%%%%%%%%%%%%%%%%%%%%%%%%%%%%%%%%%%%%%%%%%%%%%%%%%%%%%
\vspace{0.2in}
\begin{figure}[ht]
\begin{center}
\includegraphics[width=2.5in,angle=0,clip=false]{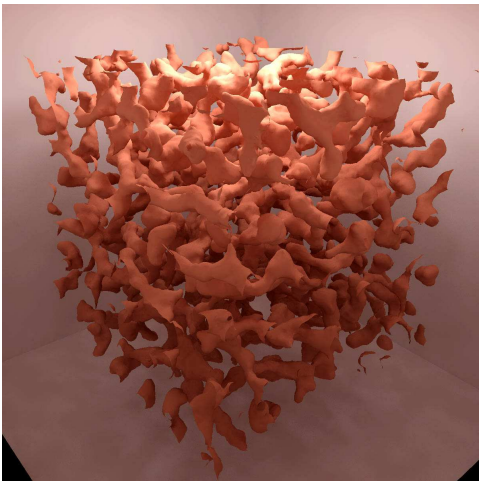}
\caption{(Color online) The 0.03~fm$^{-3}$ proton density 
          isosurface for one configuration of 100,000 nucleons 
          at a baryon density of 0.05 fm$^{-3}$. The simulation 
	  volume is a cube of 126 fm on a side.}
\label{Fig1}
\end{center}
\end{figure}
%%%%%%%%%%%%%%%%%%%%%%%%%%%%%%%%%%%%%%%%%%%%%%%%%%%%%%%%%%%%%%%%%

\begin{figure}[ht]
\begin{center}
\includegraphics[width=2.5in,angle=270,clip=false]{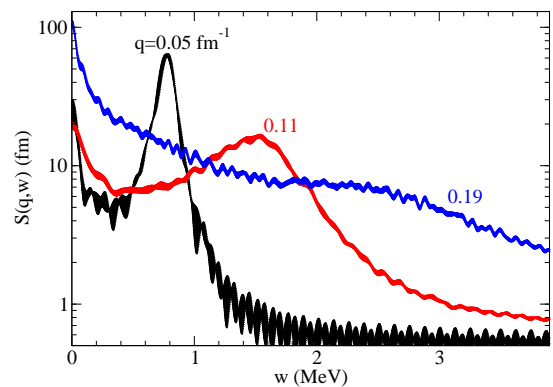}
\caption{(Color online) The dynamical response function 
          $S(q,\omega)$ versus excitation energy $\omega$ 
          at a density of $\rho\!=\!0.05$ fm$^{-3}$ and 
	  momentum transfers of $q\!=\!0.05$, 0.11, and 
	  0.19 fm$^{-1}$.}
\label{Fig2}
\end{center}
\end{figure}

At any time during the simulation one may examine the spatial
correlations of the nuclear pasta. A typical proton density is
displayed in Fig.~\ref{Fig1} at a density of
$\rho\!=\!0.05$~fm$^{-3}$.  Although protons are seen to cluster into
complex elongated shapes, it is difficult to discern a single
underlying structure ({\it e.g.,} spheres, cylinders, {\it
etc.}). Although not shown, most of the neutrons also cluster into
these complex shapes. However, in addition, there is a low-density
neutron gas between the clusters. The resulting dynamical response of
the nuclear pasta is shown in Fig~\ref{Fig2}. The choice of $T_{\rm
max}$ in Eq.~(\ref{sqw}) involves a tradeoff. $T_{\rm max}$ should be
large enough to avoid truncation errors and small enough to minimize
statistical errors from $S(q,t)$ at large $t$. A $T_{\rm max}$ of the
order of 20,000 fm/c was used.  We note that at low momentum transfers 
the dynamical response shows a peak just below $\omega\!=\!1$ MeV. This
peak becomes broader with increasing $q$. Further, there is
substantial strength near $\omega\!=\!0$.  To interpret these peaks we
calculate $S(q,\omega)$ at a lower density and compare it to the
corresponding response of a simplified ion model~\cite{pasta2}.

%%%%%%%%%%%%%%%%%%%%%%%%%%%%%%%%%%%%%%%%%%%%%%%%%%%%%%%%%%%%%%%%%
\vspace{0.02in}
%%%%%%%%%%%%%%%%%%%%%%%%%%%%%%%%%%%%%%%%%%%%%%%%%%%%%%%%%%%%%%%%%

A simulation at the lower density of $\rho\!=\!0.01$~fm$^{-3}$ (with
$Y_p\!=\!0.2$ and $T\!=\!1$ MeV) reveals nucleons clustered into more
conventional neutron-rich nuclei rather than in complex pasta shapes.
At this density, thermalization is slower because of the Coulomb
barrier; it may take a long time to add protons to a
cluster. Therefore, a system of 40,000 nucleons initially distributed
at random, had to be evolved for the very long time of 1,287,000
fm/c. During this time the temperature of the system was first raised
and then lowered to the target temperature of $T\!=\!1$~MeV. Even so,
we can not rule out a further increase in cluster size from evolution
on even longer time scales.  To compute the dynamical response, the system was evolved for another 720,000 fm/c while writing configurations to disk every 20 fm/c, for a total of 36,000 configurations.

%%%%%%%%%%%%%%%%%%%%%%%%%%%%%%%%%%%%%%%%%%%%%%%%%%%%%%%%%%%%%%%%%
\vspace{0.02in}
\begin{figure}[ht]
\begin{center}
\includegraphics[width=2.5in,angle=270,clip=false]{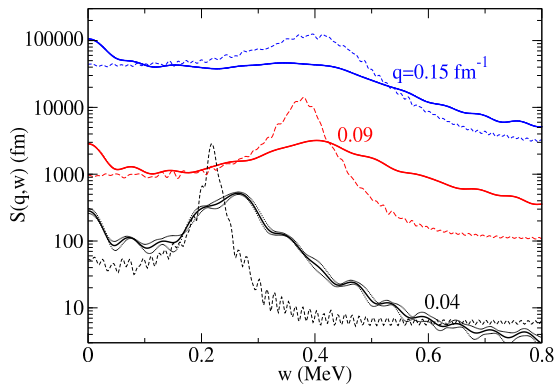}
\caption{(Color online) The dynamical response function 
          $S(q,\omega)$ versus excitation energy $\omega$ 
          at a density of $\rho\!=\!0.01$ fm$^{-3}$ and 
	  momentum transfers of $q\!=\!0.04$, 0.09, and 
	  0.15~fm$^{-1}$. The solid lines show results 
          for a 40,000 nucleon simulation while the 
          dashed lines are results for an ion model
          [see Eq. (\ref{smodel})] simulation with 
	  288 ions. For clarity the $q\!=\!0.09$ fm$^{-1}$ 
	  results have been multiplied by 10 while
          the $q\!=\!0.15$ fm$^{-1}$ results by 100.}
\label{Fig3}
\end{center}
\end{figure}
%%%%%%%%%%%%%%%%%%%%%%%%%%%%%%%%%%%%%%%%%%%%%%%%%%%%%%%%%%%%%%%%%

%%%%%%%%%%%%%%%%%%%%%%%%%%%%%%%%%%%%%%%%%%%%%%%%%%%%%%%%%%%%%%%%%%%%%%%
      
In Fig.~\ref{Fig3} the response of the nuclear pasta is compared 
to that of an ion model where the composition of the system is
assumed to be 28\% of free neutrons and $X_h=72$\% of a single 
representative heavy ion of average mass $A\!=\!100$ and charge 
$Z\!=\!28$. These numbers were obtained from counting nucleons in 
the different clusters using a procedure described in 
Ref.~\cite{pasta2}. The dynamical response is modeled as,
%%%
\begin{equation}
 S_{\rm model}(q,\omega)= X_h \langle N\rangle 
                          F(q)^2 S_{ion}(q,\omega)\;.
 \label{smodel}
\end{equation}
%%%
Here $S_{\rm ion}(q,\omega)$ is the response of $Z\!=\!28$ ions with
only screened Coulomb interactions, $\langle N\rangle\!=\!72$ is the
average number of neutrons in a cluster, and $F(q)$ is the cluster
form factor, that is approximated as a uniform density sphere of
radius 6.68 fm and normalized to $F(0)\!=\!1$~\cite{pasta2}.  $S_{\rm
ion}(q,\omega)$ is computed from a molecular dynamics simulation for a
system of 288 ions in the same volume as the 40,000-nucleon
simulation. The model response in Fig.~\ref{Fig3} shows a single peak
from plasma oscillations with a width that increases with $q$. This
peak is also visible in the full nucleon response at low $q$ and as a broad shoulder at $q\!=\!0.15$ fm$^{-1}$. The location of the peak is approximately described by the known expression for the plasma frequency~\cite{FW}:
\begin{equation}
\omega_{\rm pl} \approx 
 (4\pi Z^2e^2\rho_i/M_i)^{1/2}
 (1+(q\lambda)^{-2})^{-1/2},
\label{plasma_freq}
\end{equation}
where $\rho_i$ is the ion density and $M_i$ the ion mass.  The second
factor appearing in the above expression is a crude RPA estimate of
the decrease in the plasma frequency due to the screening length
$\lambda$ appearing in Eq.~(\ref{v}). However, note that the width of
the plasma oscillation in the nucleon simulation is much larger than
the width in the ion model. This may reflect a coupling of the plasma
oscillation to the internal degrees of freedom of the clusters and/or
failure of the single heavy-ion approximation. 

In addition, the full nucleon simulation shows a peak at $\omega\!=\!0$ that is absent from the ion model.  We speculate that this mode may be associated with density fluctuations.  A liquid vapor coexistence region has large density fluctuations as vapor is converted to or from liquid.  Note that the energy associated with transferring a neutron from the vapor to a cluster is zero because the vapor is in equilibrium with the condensed phase.  Therefore, the system can support a density wave with low excitation energy.  In the nuclear pasta --- a system with two conserved quantities (baryon number and electromagnetic charge) --- these fluctuations are constrained at long wavelengths by charge neutrality.  However, the system may still experience density fluctuations at {\it finite} $q$ as nucleons condense and evaporate from the individual clusters.   This density wave may represent a hallmark of frustrated, multicomponent systems having more than one conserved charge.  Therefore, it is important to verify our semiclassical results with full quantum calculations.  The interpretation of this $\omega\!=\!0$ mode as a density wave should also be verified in future work.  We believe our speculation is reasonable but this needs to be checked.

Finally, we compare our results at $\rho=0.05$ and 0.01 fm$^{-3}$.  It is difficult to apply our ion model directly at a density of $\rho=0.05$ fm$^{-3}$ because the masses and charges of the interconnected clusters are not well defined.  The full simulation results at $\rho=0.05$ have a higher frequency plasma oscillation compared to $\rho=0.01$.  Note, although the plasma frequency depends on density, it may not be strongly modified by the non-spherical cluster shapes that are present at high density.  The low omega mode, which is present at $\rho=0.01$, is more pronounced at $\rho=0.05$.

Between densities of 0.01 and 0.05 fm$^{-3}$, non-spherical shapes appear.  It is natural to ask how these shapes impact the excitation spectrum.  Plasma oscillations involve long distance classical physics.  The electrostatic restoring force depends only on the charge density.  Therefore the plasma frequency depends on the ratio of the ion charge density to mass as indicated in Eq. (\ref{plasma_freq}).  This ratio is independent of the shape of the clusters.  Extended shapes such as long rods could also have nuclear contributions to the restoring force.  This could raise the oscillation frequency.  However between 0.01 and 0.05 fm$^{-3}$ we find the ratio of the frequency of the $q=0.04$ fm$^{-1}$ peak in Fig. 3 to the $q=0.05$ fm$^{-1}$ peak in Fig. 2 to be in good agreement with Eq. (\ref{plasma_freq}).  This suggests that the nuclear contributions to the restoring force at a density of 0.05 fm$^{-3}$ are small.  Thus the plasma oscillations we find in the pasta simulation are consistent with an electrostatic restoring force only. 

In conclusion, we have modeled the complex nuclear-pasta phase via a
semi-classical model that reproduces nuclear saturation and includes
Coulomb repulsion. The dynamical response of the nuclear pasta is
computed from molecular dynamics simulations with 40,000 and 100,000
nucleons. We find that the nuclear pasta supports a plasma oscillation
with a frequency in the 1-to-2 MeV range.  In addition, the dynamical
response displays a substantial amount of strength at low energies,
that we identify as a coherent density wave involving the internal
degrees of freedom of the clusters. 

\smallskip
We acknowledge useful discussions with S. Reddy, D. Bossev, and
H. Kaiser. We thank Brad Futch for preparing Fig.~\ref{Fig1}.  
This work was supported in part by DOE grants DE-FG02-87ER40365 and
DE-FG05-92ER40750, and by Shared University Research grants from IBM,
Inc. to Indiana University.


\begin{thebibliography}{99}
\bibitem{footnote} Other runs where the velocities were periodically rescaled seemed to yield identical results for $S(q,\omega)$.
\end{thebibliography}
\end{document}